# A Gaze Data-based Comparative Study to Build a Trustworthy Human-AI Collaboration in Crash Anticipation

Li Y.,[1] Karim M.M.,[2] and Qin R., Ph.D., A.M.ASCE [3]

[1]Ph.D. student, Department of Civil Engineering, Stony Brook University, 100 Nicolls Rd, Stony Brook, NY 11794; e-mail: yu.li.5@stonybrook.edu
[2]Ph.D. student, Department of Civil Engineering, Stony Brook University, 100 Nicolls Rd, Stony Brook, NY 11794; e-mail: muhammadmonjur.karim@stonybrook.edu
[3]Associate Professor, Department of Civil Engineering, Stony Brook University, 100 Nicolls Rd, Stony Brook, NY 11794; e-mail: ruwen.qin@stonybrook.edu

**ABSTRACT**

Vehicles with a safety function for anticipating crashes in advance can enhance drivers' ability to avoid crashes. As dashboard cameras have become a low-cost sensor device accessible to almost every vehicle, deep neural networks for crash anticipation from a dashboard camera are receiving growing interest. However, drivers' trust in the Artificial Intelligence (AI)-enabled safety function is built on the validation of its safety enhancement toward zero deaths. This paper is motivated to establish a method that uses gaze data and corresponding measures to evaluate human drivers' ability to anticipate crashes. A laboratory experiment is designed and performed, wherein a screen-based eye tracker collects the gaze data of six volunteers while watching 100 driving videos that include both normal and crash scenarios. Statistical analyses of the experimental data show that, on average, drivers can anticipate a crash up to 2.61 seconds before it occurs in this pilot study. The chance that drivers have successfully anticipated crashes before they occur is 92.8%. A state-of-the-art AI model can anticipate crashes 1.02 seconds earlier than drivers on average. The study finds that crash-involving traffic agents in the driving videos can vary drivers' instant attention level, average attention level, and spatial attention distribution. This finding supports the development of a spatial-temporal attention mechanism for AI models to strengthen their ability to anticipate crashes. Results from the comparison also suggest the development of collaborative intelligence that keeps human-in-the-loop of AI models to further enhance the reliability of AI-enabled safety functions.

**INTRODUCTION**

Safety enhancement is a priority of transportation. In 2019, there were 6,756,000 crashes in the U.S., and 33,244 of these were fatal crashes (NHTSA 2020a). 36,096 people were killed in the crashes, and 2740,000 people were injured. The consequence of crashes is huge. The average comprehensive cost is $11.1 million per death and $1.2 million per disabling (National Safety Council, n.d.). The major factor in 94% of all fatal crashes is human error. Technologies that can enhance roadway safety are receiving growing attention. Automated Driving Systems (ADSs) have a promise to help drivers avoid crashes (NHTSA 2020b).

The crash anticipation function of ADSs has been an important research problem for computer vision and deep learning that develop deep neural networks to analyze dashboard camera (dashcam) captured videos for predicting crashes in advance (Chan et al. 2016; Bao et al. 2020; Li et al. 2021; Karim et al. 2022a). AI models that can perform better than or comparably to humans, and those whose decisions are explainable to humans (Karim et al. 2022b), are likely to gain trust from ADS users and the public (Xu et al. 2019). AI models were also developed to predict humans' visual attention (Deng et al. 2016; Deng et al. 2019; Xia et al. 2018), which are used for a wider range of safety applications such as saliency detection, hazard anticipation, and attention allocation to name a few. AI research on crash anticipation and safety enhancement is developing rapidly. Yet, how well humans anticipate crashes visually is less known.

Although existing AI models that have methods, measures, and metrics for evaluating the performance of crash anticipation, those are not applicable to human drivers. Drivers' anticipation of a crash is a subjective judgment. Precisely measuring their ability to anticipate crashes is challenging. For example, the earliness of an AI model in anticipating a crash is measured by the Time-To-Crash (TTC), from the earliest time when the AI model's prediction score has reached a threshold to the starting time of the crash. The TTC for a human is difficult to measure. The self-estimation of TTC by drivers, for example using a survey method, is not precise and may have biases. It is difficult for a driver to tell exactly when she/he has been aware of the risk and anticipated a crash. The performance of an AI model in classifying a driving scene as normal or risky without seeing an accident can be measured by metrics such as recall and precision. Yet, these are not directly applicable to humans because of the difficulty in measuring humans' judgment precisely. Self-reporting methods usually are time-consuming.

An eye tracker is a sensor that can sense and measure humans' gaze points, pupil diameters, and eye positions with high accuracy. It has been used widely for studying diagnostic systems by providing objective and quantitative evidence of users' visual attention or serving as an input device of interactive systems in a host of visually mediated applications (Duchowski 2002). For example, eye trackers can extract humans' fixation points from their gaze points. Fixation points of a person can tell what she/he attended to, when, and how long, thus revealing the person's thinking, reasoning, responding, and judgment. Metrics created based on humans' fixations may provide accurate measurements for inferring drivers' ability to anticipate crashes.

Facing the need for measuring humans' ability to anticipate traffic crashes and the lack of methods, measures, and metrics for the research, this paper is motivated to perform a gaze data-based pilot study, which aims to reduce the gap to the rapid advancement of AI research. Contributions of this paper are threefold:
- gaze data-based methods, measures, and metrics for quantifying drivers' ability to anticipate traffic crashes in advance,
- a laboratory experimental study whereby gaze data were collected and analyzed for inferring drivers' crash anticipation ability,
- a comparison between an AI model and humans on their crash anticipation performance.

The remainder of the paper is organized as the following. The next section reviews the related literature to determine the research status. Then, the proposed methods, measures, and metrics for assessing drivers' ability to anticipate traffic crashes are presented. After that, results and findings from an experimental study are discussed. In the end, the paper concludes the study and proposes future work.

## LITERATURE REVIEW

Eye trackers have been adopted by transportation safety studies for at least two decades. In the early stage, it was mainly used for understanding drivers' behavior. For example, the situational awareness of drivers increases in higher levels of automation, which is more likely to happen in scenarios that encourage drivers to gaze at the road center (Louw and Merat 2017). Geometric and illumination conditions of highways are found to influence drivers' visual attention (Suh et al. 2006). Drivers often concentrated on the end of the road in front of their vehicles (Deng et al. 2016). Gaze data are also used for detecting drowsiness (Rumagit et al. 2017), lane change (Jang et al. 2014; Doshi and Trivedi 2009), and hazard perception (Alberti et al. 2012; Fisher et al. 2007).

     The glance-monitoring technology enhanced the theoretical understanding of drivers' behavior and, thus, it can be further used for roadway safety enhancement (Taylor et al. 2013). The crash anticipation function of ADS is advantageous for avoiding crashes, and attention anticipation serves the purpose of crash anticipation. For example, Deng et al. (2019) proposed a human-like driving model that uses a convolutional-deconvolutional neural network to predict drivers' eye fixations by providing the most relevant regions or targets, thus largely reducing the interference of irrelevant scene information. Xia et al. (2018) built a driver attention prediction model that uses driver eye movements to identify more crucial driving moments and weight them more during model training. More attention prediction models with neural networks were proposed, and applications of those neural networks implicitly assume that humans' visual attention is highly reliable and effective. This assumption needs to be verified.

     Inspired by humans' visual attention behavior, a stream of deep learning research has integrated human-like attention models in crash anticipation neural networks. Chan et al. (2016) and Bao et al. (2020) learned a spatial attention model that weighed different traffic agents for crash anticipation. Karim et al. (2022a) developed a dynamic spatial-temporal attention model that largely improves the crash anticipation performance. Unlike these attention models learnt from crash video data, Li et al. (2021) developed a spatio-temporal attention mechanism from big crash report data, which supplements the attention learnt from crash video data by extending the earliness of risk perception. Although those neural networks for crash anticipation have good performance, they are like a black box for drivers and the public. AI models that are explainable to humans are more likely to gain the trust from users and be accepted by them (Xu et al. 2019; Karim et al. 2022b). An intuitive method to improve the explainability of these AI models is to reveal how humans are making the same decision.

## METHODS, MEASURES, AND METRICS

This section introduces the data collection methods, as well as the measures and metrics developed based the corrected data, for studying drivers' ability in anticipating traffic crashes. Table 1 below first lists the symbols used in this paper.

**Table 1. List of Symbols.**

| Symbols | Meaning |
|---|---|
| $CIO$ | Crash-involving Objects |
| $D$ | Early Attention Duration |
| $F_k$ | Fixation |

| $L$ | Latency |
| --- | --- |
| $R_H$ | The cap of humans' recall in crash anticipation |
| $TTC$ | Time-to-Crash |
| $mD$ | Video-level mean early attention duration |
| $mTTC$ | Video-level mean time-to-crash |
| $\rho_R(D)$ | Percent fixation duration in the early attention duration |
| $\rho_F(T_B \cup L)$ | Percent fixation duration before the driver fixates on CIOs |
| $\rho_R(D)$ | Percent fixation duration on CIOs in the early attention duration |

**Gaze Data Collection.** This study collected drivers' gaze data by letting participants watch dashcam captured driving videos in a lab setting. This approach has been proved to be valid and it has some advantages over collecting the data in the field study (Doshi and Trivedi 2009). 100 videos of diverse driving scenes were sampled from the CCD dataset (Bao et al. 2020). 50 of these are positive videos that each contains a crash, and the remaining 50 are negative videos that have no crash at all. Each video lasts 5 seconds and the frequency of the videos is 10 Hz. That is, each video is a sequence of 50 frames of images. The starting time of the crash in each positive video is random, ranging from 3 to 5 seconds. The sequence of the 100 videos were randomized so that subjects did not know the class of the next video. A one-second interval of a blank image is placed into any two adjacent videos so that participants have a chance to rest for a short period before transitioning to the next video. Therefore, the entire video sequence for collecting human gaze points lasts 10 minutes.

Tobii Pro Fusion is a screen-based eye tracker used in this pilot study for collecting gaze points. The eye tracker provides the coordinates of each gaze point on a video frame and the timestamp. The frequency of the eye tracker is 120 Hz, which allows for collecting up to 12 gaze points per frame from each participant, approximately. According to the angular speed of their eye movements, gaze points are classified as fixation points, saccade points, or unknown. The time series of fixation points of a driver can tell what attracts the driver's attention, when, and how long. A group of successive fixation points of a person is named a fixation. Let $k$ be the index of a participant's sequential fixations in watching a video. A fixation is described as:

$$F_k = \{s_k, S_k, C_k\}, \tag{1}$$

where $s_k$ is the starting time of the fixation, $S_k$ is its duration, and $C_k$ is its coordinates calculated as the centroid of gaze points belonging to this fixation. Regions of a frame where drivers fixate on are those attracting their attention.

This pilot study invited six volunteers to participate in experiment to capture inter-subject variability. Their ages are between 21~45 years old. All participants have a driver's license with 2~18 years of driving experience. Each volunteer watched the video sequence twice to capture inter-subject variability. They took a break for at least 30 minutes between the two times of experiments. In total, the study collected 720,000 gaze points from this study, approximately, about 144 gaze points per frame. The size is comparable to those in literature.

**Measuring the Earliness of Crash Anticipation.** While metrics for measuring the earliness of AI models in anticipating future crashes are well established, those for humans do not exist. This study developed metrics for measuring the earliness of drivers in anticipating traffic crashes. Figure 1 illustrates the sequential events associated with a positive video. The earliest time when a Crash-Involving Object (CIO) appears in the video splits each positive video into two stages: the duration before any CIO appears, $T_B$, and the duration with CIOs, $T_A$. After a CIO appears, the driver may not attend to it immediately. Latency, $L$, is defined as the duration from the earliest time when a CIO appears to the first time when the driver's fixation point falls on a CIO. If there are multiple CIOs, the time when the earliest appeared CIO shown in the video is taken as the start of the latency. A short latency indicates a better ability to anticipate crashes earlier.

Time-To-Cash ($TTC$) is defined as the time period from when a crash has been anticipated to the start of the crash. The longer the $TTC$ value, the earlier the crash anticipation. Measuring the $TTC$ for drivers is difficult because their anticipation of a crash is a subjective judgment. However, it is safe to claim that the driver anticipates the occurrence of a crash somewhen between the earliest time a CIO captures the driver's attention and the start of the crash. This time period is termed early attention duration, $D$. During this period, the driver's attention on CIOs accumulates, and so the driver's perception of the crash risk is developing. $D$ is the upper boundary of $TTC$ (i.e., $D \geq TTC$) and $D$ can be measured accurately by an eye tracker. Therefore, this study chose to use the early attention duration to infer drivers' $TTC$. $D$ varies among drivers and driving scenes. $mD$ denotes the video-level mean value of different participants' $D$ values.

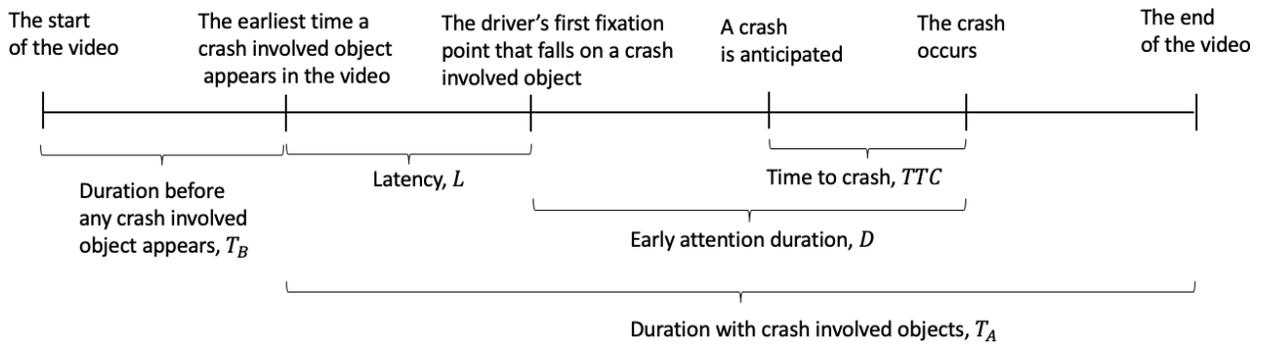

**Figure 1: Schematic diagram for defining the measures and metrics for the earliness in crash anticipation.**

The mean $TTC$ denoted as $mTTC$ in this study is used to present the earliness of AI model developed by Karim et al. (2022) recently to represent crash anticipation models because it establishes a new state-of-the-art.

**Drivers' Early Attention Level on Crash Involved Objects**. Figure 1 illustrates that CIOs have caught the driver's attention at the beginning of the early attention duration $D$. The more attention that CIOs receive during $D$, the larger of the likelihood that the driver has perceived the crash risk. This study annotated the CIOs in the 50 positive videos using the VGG annotator (Dutta et al. 2016). $B_{t,n} = [x_{t,n,1}, x_{t,n,2}, y_{t,n,1} \ y_{t,n,2}]$ denotes the bounding box of the $n^{th}$ CIO in frame $t$ of a video, where $(x_{t,n,1}, y_{t,n,1})$ are the coordinates of the lower left point and $(x_{t,n,2}, y_{t,n,2})$ are the coordinates of the upper right point.

Since not all gaze points are fixation points, fixations may just be part of any time period. For example, the driver's cumulative fixation duration, as a fraction of the early attention duration, is calculated as

$$\rho_F(D) = \frac{(\cup_k [s_k, s_k + S_k]) \cap [T_B+L, T_B+L+D]}{D}, \quad (2)$$

which measures the average attention level of the driver during the early attention duration $D$. The study also measures the driver's cumulative fixation duration on CIOs, as a fraction of the early attention duration:

$$\rho_R(D) = \frac{(\cup_k [s_k, s_k + S_k]) \cdot 1\{C_k \in \cup_{t,n} B_{t,n}\}) \cap [T_B+L, T_B+L+D]}{D}, \quad (3)$$

which is the driver's average attention level on CIOs during $D$. The ratio $\rho_R(D)/\rho_F(D)$ measures the fraction of the driver's attention allocated to CIOs during $D$. The higher this ratio is, the larger the likelihood whereby the driver has perceived the crash risk and thus anticipated a crash in advance.

**Reliability in Crash Anticipation.** Due to the difficulty in evaluating the drivers' subjective judgment of video classes, this study is not able to precisely measure the recall and precision of a driver in anticipating crashes from the sample of videos. Instead, their upper bounds are estimated based on the driver's fixation points. On a positive video, if a driver never fixated on CIOs before the crash occurs, the driver failed to perceive the risk in advance. This driver's classification result on this positive video must be a false negative. However, having fixations on CIOs before the crash occurs is a necessary condition, but not a sufficient condition for a driver to develop the early perception of crash risk. Therefore, the study defines

$$R_H = \frac{\text{\#positive videos wherein CIOs receive attention before the crashes occur}}{\text{\#positive videos}} \quad (4)$$

as the upper bound of a driver's recall. Drivers' false positive rate usually is low, and the consequence of a false positive classification is way less severe than that of a false positive. Therefore, this study assumes drivers' precision is assumed 100% in this study.

**RESULT ANALYSES**

With the methods, measures, and metrics developed in this study, humans' ability to anticipate traffic crashes is analyzed and compared to the AI model. The analysis is focused on four aspects: the temporal and spatial dynamics of drivers' attention, drivers' earliness in crash anticipation, drivers' attention level on CIOs, and reliability of drivers' crash anticipation. The level of significance, $\alpha$, is 0.05 in all the statistical analyses.

**Temporal Dynamics of Drivers' Instant Attention Level**. A portion of a driver's gaze point on each frame is fixation points. The driver's fixation points on each frame in percent is a proxy of the driver's instant attention level. Figure 2 summarizes participants' instant attention level in the experimental study. Results pertaining to the positive videos are in the first row, and those with the negative videos are displayed in the second row. The two time series plots in the first column are individual subjects' instant attention level. Overall, a driver's instant attention level is dynamic, varying from one frame to another frame and from one video to another video. Drivers' instant attention level on negative videos is more stable along the timeline than that on positive videos. Plots in columns 2 and 3 are examples of instant attention time series. The examples indicate that the participant's instant attention level could be varied largely by the occurrence of crashes. Column 4 further shows the distributions of the instant attention level. The two distributions have different shapes, central locations, and ranges, indicating drivers' instant attention level can be varied by driving video classes. For example, the mean instant attention level on positive videos is 0.85±0.004 and that on normal scenes is 0.82±0.004. The instant attention level on positive videos is distributed more widely (from 0.25 to 1) than that on negative videos (from 0.5 to 1).

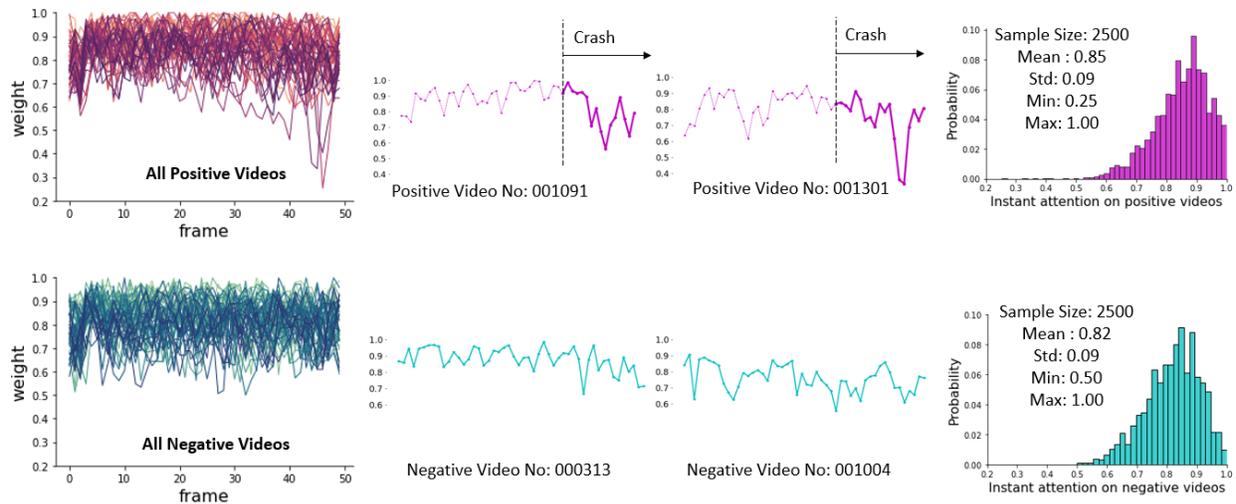

**Figure 2: Dynamics of drivers' instant attention level**

**Spatial Distribution of Drivers' Attention.** The spatial distribution of drivers' fixations represents their spatial attention. Figure 3 shows the heat maps of fixations for the two classes of videos, where colored regions are those with fixations and red-colored spots are regions with a high density of fixations. Differences between the two heat maps are noticed. Firstly, fixations on the positive videos span wider along the horizontal line than on the vertical line, whereas is not observed from the negative videos. Secondly, red spots in the heat map of the negative videos are clustered closely on a narrower area at the center of frames, whereas those of the positive videos clearly have a wider distribution. The heat map of fixations on the negative videos reveals the attention behavior of drivers in normal driving scenes. That is, drivers often fixate on the end of the road in front of their vehicles and check the surrounding traffic agents and the environment occasionally. The heat map of fixations on the positive videos indicates that drivers look at surroundings more often instead of a straight-ahead stare when they are driving in risky scenes. That is, CIOs and their motions could partially attract drivers' attention away from what they normally fixate on.

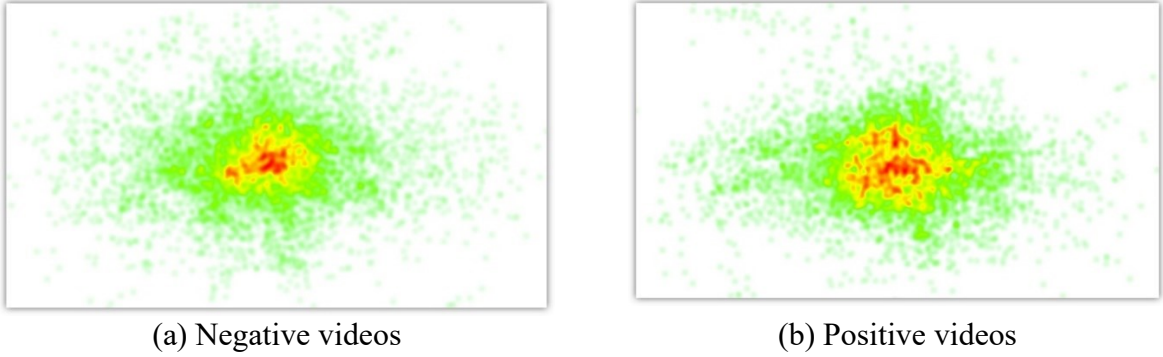

(a) Negative videos                (b) Positive videos

**Figure 3: Spatial distribution of drivers' attention: Heat map of fixation count**

**How Early Can Drivers Anticipate a Crash?** Both the early attention duration $D$ and the latency $L$ are metrics of the drivers' earliness in crash anticipation. Among the 600 experiments on the positive videos, there are 27 experiments wherein participants missed the CIOs. Figure 4 displays the distributions of the latency $L$ and the early attention duration $D$ based on the data from the remaining 573 experiments. The statistical measurements in Figure 4(a) shows that drivers' mean latency is 0.81±0.078 seconds. That is, on average, drivers notice CIOs 0.73~0.89 seconds after they appear. After CIOs show up in the video, with a 70% chance, drivers will notice the CIOs within 1 second. But the latency has a wide distribution, with a long tail skewed to the right, indicating the latency could be long for some drivers under some circumstances. The measurements in Figure 4(b) indicate the mean value of the early attention duration $D$ is 2.61±0.100 seconds. That is, on average drivers can anticipate traffic crashes up to 2.61 seconds in advance. The chance that $D$ is longer than 2 seconds is about 70%, and the chance that it is longer than 3 seconds is reduced to 35%. A negative value of $D$ means the driver did not notice the COIs before the crash occurs.

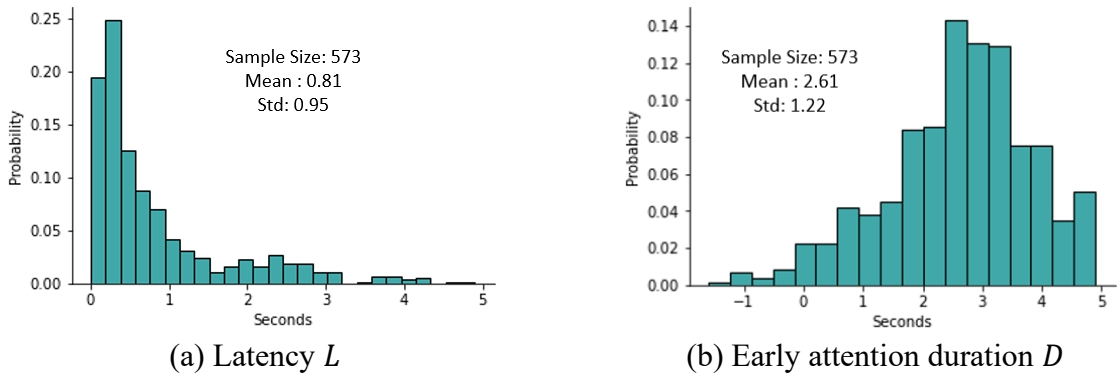

(a) Latency $L$                (b) Early attention duration $D$

**Figure 4: Distributions of the latency L and the early attention duration D**

**Does the Earliness in Crash Anticipation vary Among Drivers?** A one-way ANOVA was performed to determine if the early attention duration $D$ varies among drivers. The null hypothesis is that all participants have the same mean value of $D$. The critical F-value $F_{0.05}(6-1, 100-1)$ is 2.31. The ANOVA on $D$ shows the F-value and P-value are 3.57 (>2.31) and 0.00 (<0.05), respectively. Therefore, the null hypothesis is rejected safely, indicating at least one participant has a different mean value of $D$. Similarly, the ANOVA on latency $L$ shows F-value and P-value are 5.68 (>2.31) and 0.00 (<0.05), respectively. Again, the null hypothesis, which says all participants have the same mean value of $L$, is rejected at the level of significance 0.05, meaning that at least one participant has a different mean value of the latency than other participants. The ANOVA studies indicate drivers are heterogeneous in terms of earliness in crash anticipation.

**Whose is Earlier in Crash Anticipation, Human or AI?** Figure 5 compares drivers' early attention duration $D$ with $mTTC$ of the AI model, on each of the 50 positive videos. Each dot is the $D$ value obtained from one experiment, and 573 effective values are obtained from the 600 experiments. The green line is the video-level mean value of $D$, denoted by $mD$, and the red line is the video-level $mTTC$. Only 34 out of 573 $D$ values exceed their corresponding $mTTC$, and only 2 out of 50 $mD$ values are longer than their corresponding $mTTC$. The video-level difference between $mTTC$ and $mD$ ranges from -0.05 to 3.17 on the 50 videos. The mean difference is 1.02±0.262. Since $D$ is the upper boundary of drivers' $TTC$, the difference in anticipation earliness between AI and humans is expected to be larger than 1.02 seconds. From the analysis, it is concluded that on average the AI model is at least 1.02 seconds earlier than drivers in anticipating traffic crashes in this study.

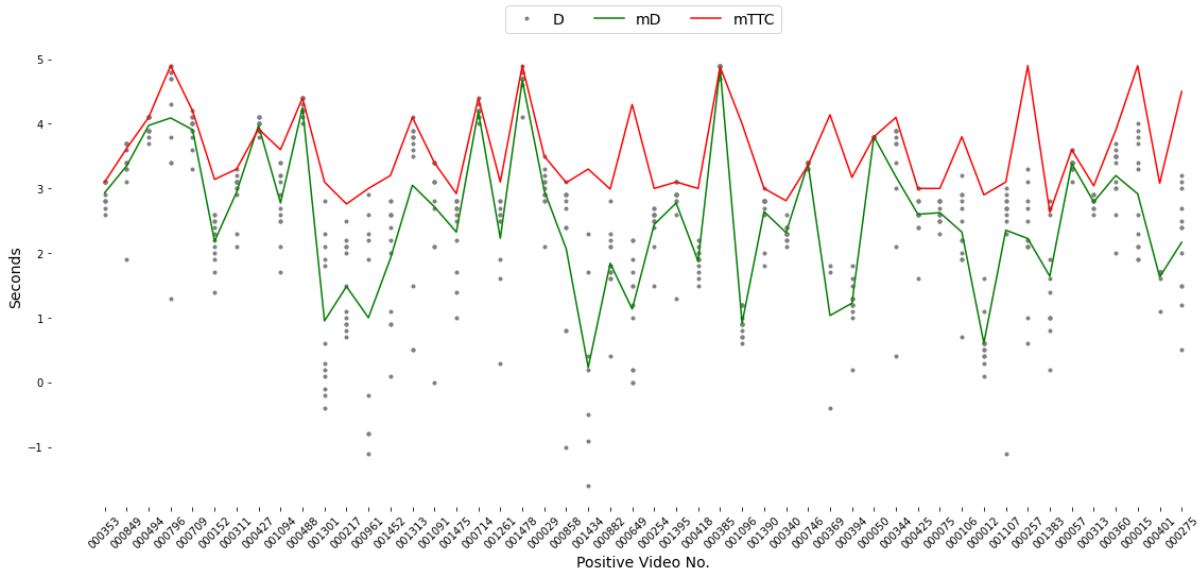

**Figure 5: Video-level $D$, $mD$, and $mTTC$**

**Drivers' Average Early Attention Level on CIOs.** Figure 6(a) shows the drivers' average attention level before fixating on CIOs, $\rho_F(T_B \cup L)$, and their average attention level during $D$, $\rho_F(D)$. Before drivers noticed CIOs, the distribution of drivers' average attention is located at 0.65 and slightly skew to the left (skewness=-1.02). During the early attention duration $D$, the location of drivers' average attention level moves to 0.83 and the distribution becomes slimmer (kurtosis=6.94) with a long tail on the left (skewness=-2.41). That is, drivers' average attention level has a clear change after they fixated on CIOs. The difference between the two distributions indicates drivers become more alert after they catch CIOs.

To determine the fraction of the average attention level during $D$ allocated to CIOs, the ratio $\rho_R(D)/\rho_F(D)$ defined in Equation (3) is calculated for each of the 560 experiments with a positive $D$ value. Figure 6(b) illustrates the distribution of the ratio $\rho_R(D)/\rho_F(D)$. The mean ratio value is 0.65±0.023. That is, on average 65% of the attention in the early attention duration is allocated to CIOs.

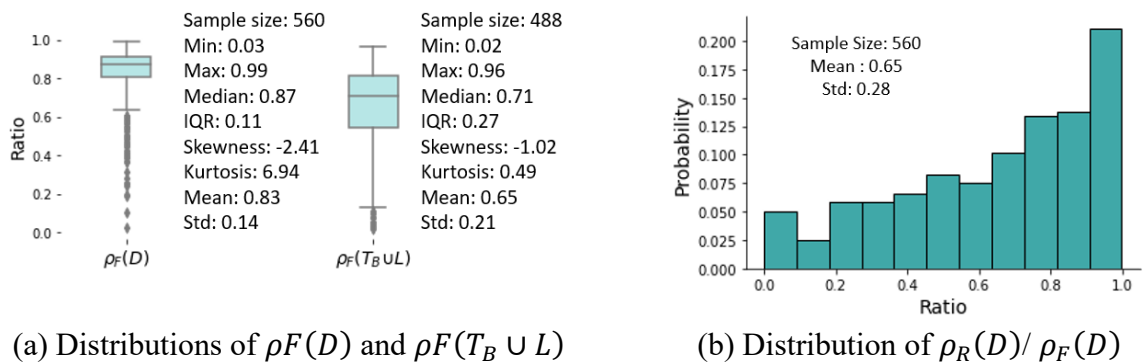

(a) Distributions of $\rho F(D)$ and $\rho F(T_B \cup L)$     (b) Distribution of $\rho_R(D)/\rho_F(D)$

**Figure 6: Average attention level in the early attention Duration, and the fraction on COIs**

**Reliability of Humans in Crash Anticipation.** Drivers' anticipation of a crash is a subjective judgment. As discussed earlier in the paper, it is difficult to precisely measure the recall and the precision of drivers in crash anticipation. In 557 out of 600 experiments, participants fixated on CIOs during the early attention period. $R_H$ in this study is 0.928±0.021. Because $R_H$ is an upper boundary of drivers' true recall value, 0.928 is the best chance that drivers may perceive the crash risk.

## CONCLUSION

This paper developed a gaze data-based method with measures and metrics for evaluating drivers' ability to anticipate traffic crashes. This establishment allows for comparing human drivers to AI models in a quantitative manner. An experimental study was designed and performed to collect gaze data from six volunteers when they were watching dashcam captured traffic videos. Statistical analysis of the experimental data shows that on average drivers can anticipate a crash up to 2.61 seconds before it occurs. The AI model can predict crashes 1.02 seconds earlier than drivers. To drivers, the AI model's earliness in crash anticipation is promising. Drivers can perceive crash risks in advance with a probability of 0.928, which means drivers themselves are highly reliable in perceive crash risks visually. The analysis also found that crash-involving objects change drivers' instant attention level, average attention level, and spatial attention distribution. The findings suggest a dynamic spatial-temporal attention mechanism should be developed and embedded to crash anticipation neural networks. Humans and AI are found to have complementary strengths, thus forming a collaborative relationship between them will improve the crash anticipation capability.

The comparison between humans and AI in crash anticipation can be further expanded. For example, to what extent do AI-proposed traffic agents overlap with those fixated by drivers?

On traffic agents that attract attention from both AI and humans, what are the attention weights assigned by AI and by humans, respectively? Besides human fixations, other data captured by the eye tracker can also be included to perform a more elaborate analysis. These include pupil diameters, the spatial leap of gaze points, and eye images data. To further explore these research questions, additional data will be collected from a larger size of subjects with more diverse backgrounds. Data will be collected both from the field and in the lab using more comprehensive designed experiments. Improving the accuracy of object detection and creating a weakly-supervised object tracking deep neural network are desired to better support the study.

**AKNOWELDGEMENT**
All authors receive support from National Science Foundation through the award ECCS-#2026357. Any opinions, findings, and conclusions or recommendations expressed in this material are those of the authors and do not necessarily reflect the views of the National Science Foundation.